\documentstyle[12pt]{article}
\input psfig.sty
\def\title#1{\relax\vspace*{2cm}{\large{\bf #1}}\par\vspace*{13.5pt}}
\def\author#1{{#1}\par\vspace*{13.5pt}}
\def\affil#1{{\it #1}\par}
\def\abstract{\vspace*{27pt}ABSTRACT\par\relax}
\def\section#1{\par{#1}\par}
\def\subsection#1{\par\underline{#1}\par}
\def\subsubsection#1{\par\underline{#1.}\ \ }
\def\acknow{\par ACKNOWLEDGMENTS\par}
\newenvironment{references}{\section{REFERENCES}\vspace*{.5cm}%
\parindent=0pt\frenchspacing%
\parskip=1pt plus 1pt minus 1pt%
\interlinepenalty=1000\tolerance=400%
\pretolerance=10000\hyphenpenalty=10000%
\everypar={\hangindent=1.6pc}
}{}

\newcommand{\Mgb}{Mg{\it b}}
\newcommand{\Mgg}{Mg$_{2}$}
\newcommand{\Cat}{Ca~II triplet}
\newcommand{\sigs}{$\sigma_{*}$}
\newcommand{\Mb}{$M_{bul}$}

\begin{document}

\title{STELLAR ABSORPTION LINES IN THE SPECTRA OF SEYFERT GALAXIES}
\author{Charles Nelson,$^1$ Mark Whittle $^2$}
\affil{$^1$ {Physics Dept. University of Nevada Las Vegas, 4505 Maryland Pkwy. 
Las Vegas, NV, 89154 USA} \\
$^2$ {Dept. of Astronomy, University of Virginia, Box 3818, Charlottesville, 
VA, 22903 USA}}

\abstract

We have measured the strengths of \Cat \ and \Mgb \ stellar absorption
lines in the nuclear and off-nuclear spectra of Seyfert
galaxies. These features are diluted to varying degrees by continuum
emission from the active nucleus and from young stars. \Cat \
strengths can be enhanced if late-type supergiant stars dominate the
near-IR light. Thus, objects with strong \Cat \ and weak \Mgb \ lines
may be objects with strong bursts of star formation. We find that for
most of our sample the line strengths are at least consistent with
dilution of a normal galaxy spectrum by a power law continuum, in
accord with the standard model for AGN. However, for several Seyferts
in our sample, it appears that dilution by a power law continuum
cannot simultaneously explain strong \Cat \ and relatively weak
\Mgb. Also, these objects occupy the region of the IRAS color-color
diagram characteristic of starburst galaxies. In these objects it
appears that the optical to near-IR emission is dominated by late-type
supergiants produced in a circumnuclear burst of star formation.

\section{INTRODUCTION}

The nuclear spectra of Seyfert galaxies can be thought in terms of two
components: a stellar component characterized by absorption lines from
stellar atmospheres and an active component composed of strong
emission lines and a featureless continuum, or FC. In the standard
model for active galactic nuclei the UV/optical continuum is produced
by non-thermal processes, following a power-law of the form $f_{\nu}
\sim \nu^{-\alpha}$ with a typical value of $\alpha = 1.0$. However,
hot, young stars also produce a featureless continuum which weakens
the absorption lines of the older stellar population.

By comparing the strengths of the \Cat \ ($\sim \lambda 8600$\AA) and
\Mgb \ ($\sim \lambda 5200$\AA) absorption lines, we can study the
diluting spectrum. An analysis of the \Cat \ features for a sample of
active and normal galaxies was carried out by Terlevich, {\it et al.}
(1990, TDT).  They concluded that the relative strengths of the \Cat \
and \Mgb \ features in Seyferts are inconsistent with dilution of a
normal galaxy spectrum by a power law continuum. They suggest that Ca
II triplet, which is known to be gravity sensitive, is enhanced by the
presence of late-type supergiants produced in a nuclear starburst.  
To determine the relative contributions of old stars, young stars and
active continuum to the spectra of Seyfert galaxies, we have done
a similar analysis using a larger sample, matched apertures for
\Cat \ and \Mgb, and a well-defined bandpasses system for the \Mgb \
line strengths.

\section{OBSERVATIONS \& PREVIOUS RESULTS}

Nelson \& Whittle (1995, NW95) obtained nuclear optical and near-IR
spectra for a large sample of Seyfert galaxies to measure their
stellar velocity dispersions, \sigs.  Sample selection was based
partly on the presence of stellar absorption features in published
spectra, resulting in a bias toward Sy~2s. We measured the absorption
line strengths in the NW95 spectra using the Burstein {\it et al.}
(1984) bandpasses for \Mgg \ and the TDT bandpasses for \Cat. A
modficiation to the \Mgg \ bandpasses was required to avoid the [NII]
line at 5200 \AA. A transformtion was applied to place the Seyferts
back on the standard system and gave good results in tests for sample
galaxies with published line strengths. \Cat is sometimes found in
emission in Seyferts but this occured for only one galaxy in the
sample (see NW95).  Nelson \& Whittle (1996), used \sigs \ to study
the dynamics of Seyfert bulges. They find that the \sigs~--~\Mb \
(Faber-Jackson) relation for Seyferts is offset from that for normals
in the direction of lower $M/L$ suggesting a significantly younger
stellar population.

\begin{figure}
\vskip -0.70cm
\hskip 4.6cm
\psfig{figure=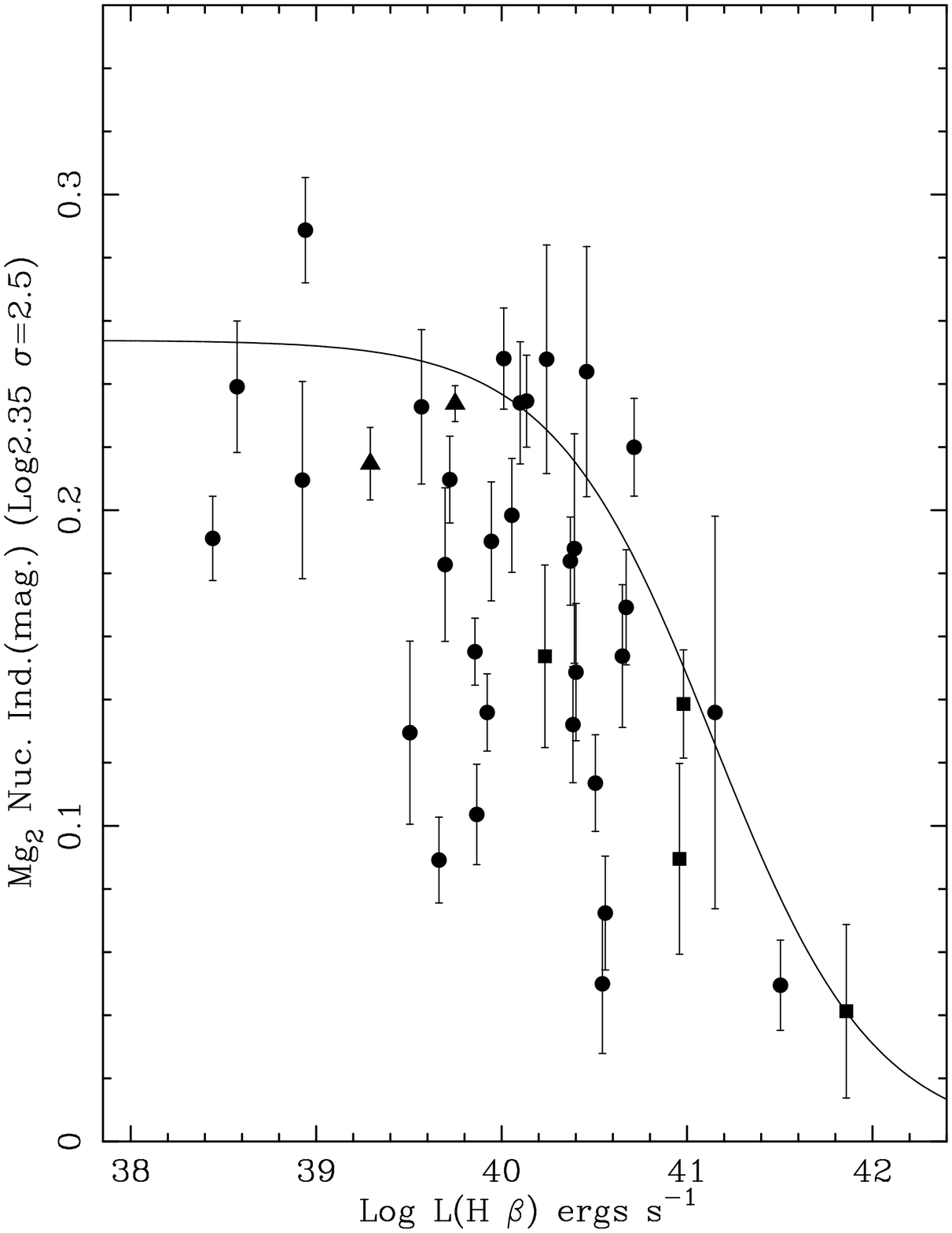,height=12cm}
\vskip -0.2cm Fig. 1. The luminosity of the H$\beta$ emission line is
plotted against \Mgg \ index corrected for the slope of the \Mgg \ --
\sigs \ relation.  Dilution of a typical galaxy by a FC following the
$L_{H\beta}$ - $L_{FC}$ relation of Yee (1980) is shown as the solid
line. Note the tendency for many Seyferts to weaker \Mgg \ than
expected.
\end{figure}

\section{\Mgg \ AND H$\beta$ LUMINOSITY}

An important relationship exists between \Mgg \ and \sigs \ in
elliptical galaxies and spiral bulges (Bender, {\it et al.}  1992,
Jablonka, {\it et al.} 1996). Using \sigs \ from NW95, we find that
the \Mgg \ -- \sigs \ relation is weak in Seyferts (Pearson linear
correlation coefficient, $R_P=0.16$), which tend to have weaker lines
than normal galaxies of the same \sigs. However, the correlation
strengthens considerably when H$\beta$ luminosity is included as a
third parameter (at fixed $L_{H\beta}$, $R_P\vert_{L_{H\beta}}=0.44,
P(null)=0.9$\%). Furthermore, the slope of original \Mgg--\sigs \
relation is recovered. The $L_{H\beta}$ dependence can be shown by
first removing the dependence on \sigs. In Figure 1, we plot corrected
\Mgg \ vs. $L_{H\beta}$ revealing a moderately strong anti-correlation
($R_P=-0.57, P(null)=0.03$\%). One possibility is that this results from
the correlation between $L_{H\beta}$ and FC luminosity for luminous
AGN (Yee 1980, Shuder, 1981). If so we can calculate the expected \Mgg
\ assuming dilution of an old stellar population using Yee's relation
between $L_{H\beta}$ and $L_{FC}$ (curving line in Figure 1).
Interestingly, in many Seyferts the \Mgg \ values are weaker than
expected, suggesting that the $L_{H\beta} - L_{FC}$ relation is not
appropriate for Sy 2 galaxies in which the continuum is not viewed
directly. Other possibilities are that the \Mgg \ -- \sigs \ relation
is different for Seyferts, or that an additional source of FC
exits. These latter alternatives are suggestive of circumnuclear star
formation or younger stellar populations.

\begin{figure}
\vskip -0.95cm
\hskip -0.5cm
\psfig{figure=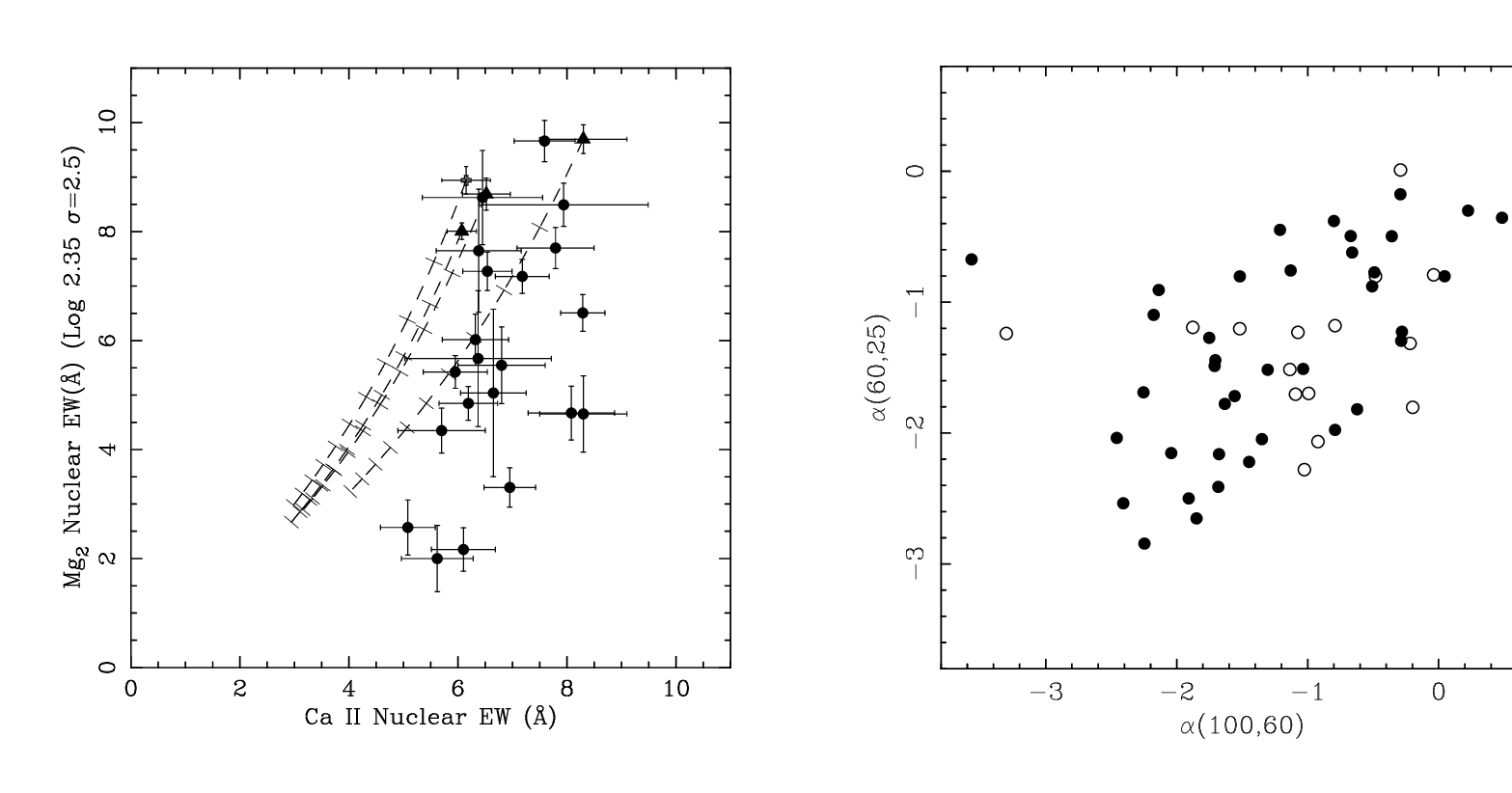,height=9.5cm}

Fig. 2. The \Cat \ and \Mgg \ line strengths are plotted. The dashed lines
show the tracks expected for an old stellar population diluted by power-law
with $\alpha=-1.0$.

\medskip
Fig. 3 IRAS Color-Color diagram. Open symbols are Seyferts with
absorption line evidence for star formation. They tend to fall in the
region to the lower right, amidst starburst galaxies. The filled
symbols are normal Seyferts and lie on the upper left side, the region
occupied by pure AGN.

\end{figure}

We have also compared the on and off-nuclear \Mgg \ line strengths in
Seyferts.  We would expect galaxies diluted by a point source, i.e. an
AGN, to have stronger off-nuclear line strengths.  This is the case
for much of the sample.  For some galaxies however, the line strengths
on and off-nucleus are weak and equal, suggesting dilution by the same
source, which must be several hundred pc across.  Thus circumnuclear
star formation may produce the FC in many Seyferts.

\section{COMPARISON OF CA II TRIPLET AND \Mgb}

In Figure 2 we plot the \Cat \ equivalent widths against those of the
corrected \Mgg. For four elliptical galaxies, representing typical
spheroid populations, dashed lines are drawn to indicate how their
line strengths would change if they were diluted by a power-law of
increasing strength. Based on these tracks we can crudely divide the
Seyferts into two groups. The first group is those points which lie
close to the dashed lines just below the ellipticals. These are
consistent with dilution by a power-law continuum. However, because
a nuclear starburst could produce similar dilution tracks we cannot
rule out dilution by a starburst or a young stellar population in
these objects.  The second group lies off the tracks in the lower right
of the diagram.  For these galaxies power law dilution of an old
stellar population does not reproduce the observed line strengths.
This suggests that the near-IR emission is not featureless but
contains strong \Cat \ absorption.  The best explanation is
that in some Seyferts red supergiants dominate the near-IR emission
strengthening the \Cat \ lines. This is consistent with a starburst
contribution to the nuclear continuum as suggested by TDT.

\section{DISCUSSION AND SUMMARY}

Table 1 categorizes our results in terms of absorption line evidence
for young stars or non-thermal power-law as the dominant continuum
source. The consistency of these indicators is good but not perfect,
that is a Seyfert with one result suggesting star formation will have
at least one other.

Since far IR-colors can roughly distinguish between dust heated by AGN
and dust heated by a starburst, it is interesting then to consider the
far-IR colors of the sample.  In Figure 3 we plot Seyferts with
absorption line evidence (having at least two of the indicators in
Table 1) for star-formation as open symbols on the IRAS color-color
diagram (see {\it e.g.}  de Grijp {\it et al.}, 1985). These tend to
lie in the lower right of the overall distribution, corresponding with the
region of the diagram occupied by starburst galaxies. The filled
symbols lie in the region occupied by pure AGN.

From these results it appears that young stellar populations and/or
circumnuclear star formation are present in many Seyferts. Furthermore
it seems that a power-law featureless continuum is also present. We
suggest that in Seyfert galaxies the power-law continuum is always
present and the strength of the starburst component varies from object
to object and may in some cases be the dominant contributor
to the continuum in the optical and near-IR. 

\begin{centering}
\begin{table}
\Large
\vspace{-0.05truein}
\centerline{\bf Table 1}
\medskip
\hspace{0.25truein}
\begin{tabular}{|c|c|}
\cline{1-2}
{\Large \bf ~ Starburst Indicators ~} & {\Large 
\bf ~ Power-Law FC Indicators ~ } \\
\cline{1-2}
\, \large 
Off \Mgg \ --  L$_{\rm H\beta}$ curve & 
\, \large On \Mgg \ -- L$_{\rm H\beta}$ curve \\
\cline{1-2}
\, \large ${\rm Mg_{2~OFF-NUC} \simeq Mg_{2~NUC}}$ & 
\, \large ${\rm Mg_{2~OFF-NUC} > Mg_{2~NUC}}$ \\
\cline{1-2}
\, \large strong \Cat, weak \Mgg & 
\, \large \Cat, \Mgg \ on FC dilution track \\
\cline{1-2}
\end{tabular}
\end{table}
\end{centering}
\bigskip

\acknow

Thanks to John MacKenty and Donna Weistrop for their advice and
support. Also thanks to Anne Kinney, Luis Ho and the UNLV Bigelow Fund
for providing financial aid to attend this meeting.

\end{document}